# Chandra Multi-wavelength Project (ChaMP).
# II. First Results of X-ray Source Properties

D.-W. Kim, B. J. Wilkes, P. J. Green, R. A. Cameron, J. J. Drake, N. R. Evans, P. Freeman,
T. J. Gaetz, H. Ghosh, F. R. Harnden, Jr., M. Karovska, V. Kashyap, P. W. Maksym, P. W. Ratzlaff,
E. M., Schlegel, J. D. Silverman, H. D. Tananbaum, and A. A. Vikhlinin

(Smithsonian Astrophysical Observatory, Cambridge, MA 02138)

August 26, 2003

Abstract

The Chandra Multi-wavelength Project (ChaMP) is a wide-area (~14 deg$^2$) survey of serendipitous Chandra X-ray sources, aiming to establish fair statistical samples covering a wide range of characteristics (such as absorbed AGNs, high z clusters of galaxies) at flux levels ($f_X \sim 10^{-15} - 10^{-14}$ erg sec$^{-1}$ cm$^{-2}$) intermediate between the Chandra Deep surveys and previous missions. We present the first results of X-ray source properties obtained from the initial sample of 62 observations. The data have been uniformly reduced and analyzed with techniques specifically developed for the ChaMP and then validated by visual examination. Utilizing only near on-axis, bright X-ray sources (to avoid problems caused by incompleteness and the Eddington bias), we derive the Log(N)-Log(S) relation in soft (0.5-2 keV) and hard (2-8 keV) energy bands. The ChaMP data are consistent with previous results of ROSAT, ASCA and Chandra deep surveys. In particular, our data nicely fill in the flux gap in the hard band between the Chandra Deep Field data and the previous ASCA data. We check whether there is any systematic difference in the source density between cluster and non-cluster fields and also search for field-to-field variations, both of which have been previously reported. We found *no significant field-to-field cosmic variation* in either test within the statistics (~1$\sigma$) across the flux levels included in our sample. In the X-ray color-color plot, most sources fall in the location characterized by photon index = 1.5-2 and $N_H$ = a few x $10^{20}$ cm$^2$, suggesting that they are typical broad-line AGNs. There also exist a considerable number of sources with peculiar X-ray colors (e.g., highly absorbed, very hard, very soft). We confirm a trend that on average the X-ray color hardens as the count rate decreases. Since the hardening is confined to the softest energy band (0.3-0.9 keV), we conclude it is most likely due to absorption. We cross-correlate the X-ray sources with other catalogs and describe their properties in terms of optical color, X-ray-to-optical luminosity ratio and X-ray colors.

1. Introduction

The launch of the Chandra X-ray Observatory has opened a new era in X-ray astronomy. With its unprecedented, sub-arcsec spatial resolution (van Speybroeck 1997), in conjunction with its high sensitivity and low background, Chandra is providing new views of the X-ray sky 10-100 times deeper than previously possible (Weisskopf et al. 2000). Indeed, the cosmic X-ray background, whose populations have long been debated because the necessary spatial resolution was lacking, is now almost (~80%) resolved into discrete sources in deep Chandra observations, e.g., the CDF-N (Chandra Deep Field-North; Brandt et al. 2001), the CDF-S (Giacconi et al. 2001). Moretti et al. (2003) has recently reported an even higher fraction (~90%). However, the nature of these sources is still somewhat unclear (e.g., Hasinger et al. 1998). An absorbed AGN population is predicted by population synthesis models (e.g. Comastri et al. 1995, Gilli et al. 1999) as the Cosmic X-ray Background is much harder (a photon index of ~1.4) than typical AGNs which have a photon index of ~1.7 (e.g. Marshall et al. 1980, Fabian & Barcons 1992). There is some observational evidence supporting the existence of red, absorbed quasars (e.g., Kim & Elvis 1999; Wilkes et al. 2002; White et al. 2003). However, the hard sources in the deep surveys appear to be a mix of various types of narrow and broad line AGNs and apparently normal galaxies with very few of the expected type 2 AGN seen. The statistical importance of these various source types requires a large sample resulting from a wider area survey such as ChaMP. Additionally, with two highly successful X-ray observatories currently in orbit (Chandra and XMM-Newton), we will soon be able to address fundamental questions such as: whether the density and luminosity of quasars are evolving in time (e.g., Miyaji et al. 2000, Cowie et al. 2003), and how clusters of galaxies form and evolve (e.g., Rosati, et al. 2002a). We will also discover whether rare, but important objects have been missing from previous studies (e.g., blank field sources discussed in Cagnoni et al. 2002).

To take full advantage of the rich dataset available in the Chandra public archive, we have initiated serendipitous X-ray source survey, the Chandra Multi-wavelength Project (ChaMP). Owing to the high spatial resolution, identification of X-ray sources is far less ambiguous than in previous missions where many counterparts were often found within typical error circles (at least ~10 times larger). Additional information and artificial selection criteria are no longer required, leaving little bias. The ChaMP, although not as deep as the CDF, covers a wide area (~ 14 $\deg^2$) and can provide an order of magnitude more sources at intermediate flux levels ($F_x \sim 10^{-14} - 10^{-15}$ erg sec$^{-1}$ cm$^{-2}$) than either the Chandra deep surveys or the previous missions (see Figure 1 in Kim et al. 2003, hereafter Paper I). An additional advantage of a wide area survey is the ability to investigate field-to-field variations of the number density of cosmic (background) sources, which may trace filaments and voids in the underlying large-scale structure or if not detected, constrain the hierarchical structure formation.

In paper I (Kim et al. 2003, submitted to ApJS), we describe our data reduction and analysis methods uniquely developed for this project and present the first catalog obtained with an initial sample of 62 Chandra observations. In this paper, we present the results of X-ray source properties by producing the Log(N)-Log(S) relation and X-ray colors, and by comparing with data at other wavelengths. In an accompanying paper (Green et al. 2003), we present the first results of deep optical follow-up observations.

This paper (II) is organized as follows. In section 2, we briefly describe our data reduction and analysis methods. In section 3, we present the number-flux relation or Log(N)-Log(S) and test its variation between different samples and from one field to another. In section 4, we discuss spectral properties of X-ray sources with X-ray colors. In section 5, we compare with data at other wavelengths. And in section 6, we present our conclusions.

2. ChaMP X-ray data

Data selection and reduction processes are described in detail in Paper I. Here we present a brief summary. We have carefully selected 137 Chandra fields which are best-suited for ChaMP science. We have applied a ChaMP-specific pipeline (called XPIPE) to uniformly reduce the Chandra data and to generate homogeneous data products. XPIPE was built mainly with CIAO (v2.3) tools (http://cxc.harvard.edu/ciao). It consists of screening bad data, correcting instrumental effects, detecting X-ray sources (by **wavdetect**) and determining source properties (including photometric, spectral, spatial, and temporal information). We have also performed a large set of simulations to quantitatively assess the source validity and positional uncertainty of **wavdetect**-detected sources (see Paper I for detailed descriptions and results). The ChaMP-selected energy bands and the definition of X-ray colors are listed in Table 1.

Table 1. Energy bands and Definition of X-ray Colors

```
Energy band selection:
    Broad (B):      0.3 - 8.0 keV
    Hard (H):       2.5 - 8.0 keV
    Soft (S):       0.3 - 2.5 keV
    Soft1 (S₁):     0.3 - 0.9 keV
    Soft2 (S₂):     0.9 - 2.5 keV

Hardness Ratio and X-ray Colors
    HR  =  (H-S) / (H+S)
    C21 = -log(S₂) + log(S₁) = log (S₁/S₂)
    C32 = -log(H)  + log(S₂) = log (S₂/H)
```

Energy band selection:
- Broad (B): $0.3 - 8.0$ keV
- Hard (H): $2.5 - 8.0$ keV
- Soft (S): $0.3 - 2.5$ keV
- Soft1 ($S_1$): $0.3 - 0.9$ keV
- Soft2 ($S_2$): $0.9 - 2.5$ keV

Hardness Ratio and X-ray Colors
$$HR = (H-S)/(H+S)$$
$$C21 = -\log(S_2) + \log(S_1) = \log(S_1/S_2)$$
$$C32 = -\log(H) + \log(S_2) = \log(S_2/H)$$

Sixty-two fields (see Table 1 in Paper I) have been completed in XPIPE processing and follow-up manual V&V. We have found 4517 sources, after excluding false sources (such as detections due to bad pixels/columns; see Table 3 in Paper I). Further excluding the target of each observation, sources at the edge of CCDs and sources affected by pile-up (in our 62 fields they all happened to be targets), we ended up with 4005 sources. 3177 sources are within CCDID=0-3 in ACIS-I observations and CCDID=6-7 in ACIS-S observations.

# 3. Log(N)–Log(S)

## 3.1. Constructing Log(N)-Log(S)

The cumulative surface number density vs. flux relation or Log(N)–Log(S) has been extensively used in understanding the nature of the cosmic X-ray background (XRB), AGN populations and their evolution (e.g., Hasinger et al. 1993). Chandra Deep surveys have now resolved almost all the XRB (Mushotzky et al. 2000; Giacconi et al. 2001; Brandt et al, 2001). However, the Chandra Deep Field surveys cover only limited sky area. The advantage of wide-area surveys such as the ChaMP is to find a large number of sources (including rare objects) at intermediate to high fluxes. In particular, the ChaMP sources nicely cover the gap between wide but shallow ROSAT/ASCA data and deep but narrow Chandra Deep Fields data. Another important advantage of the ChaMP is the ability to address any systematic field-to-field cosmic variations, for example, to confirm the reported over-density of background sources near X-ray clusters (e.g., Cappi et al. 2001).

To compare with the existing data, we have made our Log(N)-Log(S) in the soft band (0.5-2.0 keV) and in the hard band (2-8 keV), following the convention widely used in the literature. As they are similar to our S and H energy bands (see Table 1), we convert the count rates determined in our S and H bands to the flux in the above two energy bands, respectively. To calculate energy conversion factors (ECF), we assume a power-law emission model of $\Gamma_{ph} = 1.7$ for the soft band and $\Gamma_{ph} = 1.4$ for the hard band with absorption by galactic $N_H$ determined for each observation (Stark et al. 1992). Those parameters were selected to be consistent with other results (e.g., Hasinger et al. 1993; Brandt et al. 2001) for direct comparison. As described in paper I, the time-dependent QE degradation is corrected for each observation. We note that the S-band ECF varies by about 20% (for about 20 months spanning our sample) due to the QE degradation, while the H-band ECF remains almost constant.

To construct the Log(N)–Log(S), we have started with those sources of reliable quality (i.e., 3177 sources). Completeness (as well as the Eddington bias) could still affect the result considerably at the fainter end of Log(N)–Log(S). As seen in Figure 9 in Paper I, the detection probability is a function of source counts as well as off-axis distance. Therefore, we have included only sources which are (1) bright [more than 20 soft band counts for the soft band Log(N)–Log(S), and more than 20 hard band counts for the hard band Log(N)–Log(S)] and (2) near on-axis ($D_{\text{off-axis}} < 400''$ in ACIS-I observations or detected only on CCDID=7 in ACIS-S observations). These strict selection criteria ensure the completeness of our sample (> 95%), now with 707 sources and 236 sources in the S and H band, respectively. We note that we do not exclude those extreme sources which are detected only in one energy band. Utilizing the full data set would bring the Log(N)–Log(S) down to an order of magnitude lower flux level. However, that requires careful, complex correction for incompleteness (e.g., Vikhlinin et al. 1995; Kenter and Murray 2002) and we will present the detailed study with full corrections in a subsequent paper. All the X-ray sources used in this study are included in the first ChaMP catalog (Paper I).

First we have determined the sky area coverage as a function of limiting flux. The limiting flux is calculated (per chip per energy band) for a source with 20 counts in a given effective exposure time. The sky area is calculated with the geometrical area (per chip) after correcting for inter-chip gaps and for the 10% exposure threshold, which we have applied in XPIPE to remove spurious sources (see Section 3.2 in Paper I). As we are excluding sources falling at the edge of the chip, we need to correspondingly reduce the covered sky. This correction is complicated because it varies with the source size (i.e., with the off-axis angle). Instead we simply correct this effect by the fraction of sources at the edge. Typically, it is ~8% in ACIS-S and ~3% in ACIS-I.

Then, the cumulative Log(N)–Log(S) is computed by summing up the sky area-weighted source contribution:

$$N(>S) = \sum_{S_i > S} \frac{1}{\Omega_i},$$

where $N(>S)$ is the surface number density of sources with flux $> S$, $S_i$ is the flux of the $i$-th source and $\Omega_i$ is the solid angle (sky area coverage) at the flux $S_i$.

Figure 1 shows the cumulative Log(N)–Log(S) plots which were determined separately in ACIS-I Front-illuminated (FI) CCDs (CCDID= 0-3) and ACIS-S Back-illuminated (BI) CCD (CCDID = 7). The solid histogram and open circles with error bars are our unbinned and binned data, respectively. We use only unbinned data in the following analysis, but present binned data here to indicate the amount of associated errors. The results from FI (the top panel) and BI (the bottom panel) chips are remarkably similar, indicating that any systematic error in counts-to-flux conversion factors between FI and BI is minimal. In the S band (left panels), the Log(N)–Log(S) measured with ACIS-S sources is slightly higher than that of ACIS-I, but they are fully consistent within a statistical error of 1$\sigma$. In the H band (right panel), the trend is opposite. ACIS-S data points are slightly lower than those of ACIS-I. Again the difference is not statistically significant, even less significant than that in the S band. In Figure 2, we plot the Log(N)–Log(S) determined with combined FI and BI data. They are compared with previous results. In the S band, the Chandra Deep Field North (CDF-N; Brandt et al. 2001) and ROSAT (Hasinger et al. 1993) data are marked by red x's and blue filled circles, respectively. Similarly, in the H band, the CDF-N (Brandt et al. 2001) and ASCA (Ueda et al. 1998; Ogasaka et al. 1998 – converted to $\Gamma_{ph}$ = 1.4) data are marked by red x's and blue filled circles. Also plotted is the ASCA fluctuation analysis result (blue bow-tie) by Gendreau et al. (1998). In both S-band and H-band, our data are consistent with the previous results, down to 8 x $10^{-16}$ erg sec$^{-1}$ cm$^{-2}$ and 5 x $10^{-15}$ erg sec$^{-1}$ cm$^{-2}$ for the S and H band, respectively. Our H-band data fill the large flux gap between the CDF-N and ASCA data. Our H-band data point is slightly lower at the high flux end (S = $10^{-13}$ erg cm$^{-2}$ s$^{-1}$) than the ASCA data (Ueda et al. 1998). The ASCA fluctuation analysis result (Gendreau et al. 1998) appears to be consistent with our data, suggesting the same trend. However, the difference is only marginally significant (< 2$\sigma$). With a full ChaMP sample, we will be able to confirm whether there is any statistically significant difference. Although we do not plot the Chandra Deep Field-South data (CDF-S; Giacconi et al. 2001 and Rosati et al. 2002), they are consistent with the CDF-N where they overlap with our data (see below for more discussion).

In the S band, the slope starts to flatten at ~$10^{-14}$ erg sec$^{-1}$ cm$^{-2}$, as noted in the ROSAT (Hasinger et al. 1998) and XMM data (Baldi et al. 2002). Using the maximum likelihood method (Murdoch, Crawford & Jauncey 1973), we have fitted our differential Log(N)-Log(S) in the flux range of $10^{-15}$ – $10^{-13}$ erg sec$^{-1}$ cm$^{-2}$. A single power law fit is not statistically acceptable, so a broken power law model is applied as follows:

$$\frac{dN}{dS} = K\, S^{-\beta_{bright}} \qquad\qquad S > S_{break}$$

$$= \left(K\, S_{break}^{\beta_{faint}-\beta_{bright}}\right)\, S^{-\beta_{faint}} \qquad S < S_{break}$$

The best fit slopes for the differential Log(N)-Log(S) are $\beta_{bright} = 2.2 \pm 0.2$ at the bright end and $\beta_{faint} = 1.4 \pm 0.3$ at the faint end with the break energy, $S_{break} = 6\ (\pm 2) \times 10^{-15}$ erg sec$^{-1}$ cm$^{-2}$. The normalization constant $K = 2030 \pm 210$, when S is normalized by $10^{-15}$ erg cm$^{-2}$ sec$^{-1}$. They are consistent with those previously determined by the ROSAT data (Hasinger et al. 1998) and the XMM data (Baldi et al. 2002; Hasinger et al. 2001). In particular, our results are almost identical to Baldi et al.'s (2002) results obtained in a similar flux range. The CDF-N data (Brandt et al. 2001) fitted in the fainter flux range $5 \times 10^{-17}$ to $2 \times 10^{-15}$ erg sec$^{-1}$ cm$^{-2}$ resulted in the slope of the cumulative Log(N)-Log(S), $\alpha = 0.67 \pm 0.14$, which is again in good agreement with our data ($\alpha = 0.7 \pm 0.15$ for a single power-law fit in $f_x < 5 \times 10^{-15}$ erg sec$^{-1}$ cm$^{-2}$; note that $\alpha = \beta - 1$ does not hold below $S_{break}$). We have also fitted our H band Log(N)-Log(S) in the flux range $5 \times 10^{-15}$ to $2 \times 10^{-13}$ erg sec$^{-1}$ cm$^{-2}$. A single power law is statistically acceptable with the best fit $\beta = 2.1 \pm 0.1$ and the normalization constant $K = 3160 \pm 250$, when S is normalized by $10^{-15}$ erg cm$^{-2}$ sec$^{-1}$. Our result is again consistent with the XMM result ($\alpha = 1.34 \pm 0.1$) determined in the similar flux range by Baldi et al. (2002) and the CDF-N result ($\alpha = 1.0 \pm 0.3$) determined in the slightly fainter flux range ($F_X > 1.5 \times 10^{-15}$ erg sec$^{-1}$ cm$^{-2}$) by Brandt et al. (2001). A similar flattening as in the S band may exist with a break around S $\sim 10^{-14}$ erg sec$^{-1}$ cm$^{-2}$, which was also noticed in the XMM Lockman hole deep observations (Hasinger et al. 2001).

3.2. Is there field-to-field cosmic variation?

Taking advantage of a large number of observational fields, we can compare source density distributions in various sub-samples. In particular, Cappi et al. (2001) reported a considerable over-density (in the S band) in two cluster fields, when compared to the source density in non-cluster fields. This is intriguing because the over-density might be caused by the presence of clusters, indicating cosmic spatial variation of number densities by enhancing AGNs or starburst galaxies associated with the clusters or gravitationally lensed sources. To check this possibility, we have divided the 62 observations into 2 groups – those containing 1 or more previously known X-ray clusters and the rest. We have 29 fields with clusters and 33 without clusters. We have then determined the Log(N)–Log(S) separately in the fields with or without X-ray clusters and plotted them in the top and bottom panels of Figure 3. There is no statistically significant difference between the results with or without the X-ray clusters. It appears that in the S band, non-cluster fields may have a slightly higher density than cluster fields (i.e., opposite to what Cappi et al. found), but the difference is within a statistical error of 1$\sigma$. We note that most observations of cluster fields were performed in ACIS-I, while most non-cluster observations were in ACIS-S. The amount and sense of the difference is the same as that between ACIS-S and ACIS-I, as described above. In conclusion, the small difference (if any) could be due to the slight error in the calibration between ACIS-S and ACIS-I or to the presence or absence of clusters. If the presence of clusters makes the difference, the sense of the effect is opposite to the previous suggestion by Cappi et al. (2001).

To further test any field-to-field variation in cosmic background source densities, we have calculated the expected number of sources based on our Log(N)-Log(S) for a given exposure time and then compared them with the number of detected sources in each observation. As in section 3.1, we have used only sources (1) with more than 20 counts in the S or H band and (2) either within $D_{off-axis} < 400''$ for ACIS-I or in S3 for ACIS-S. In Figure 4, the ratio of detected to expected source numbers is plotted against the exposure time. Data taken from different energy bands and different CCDs are marked by different symbols and colors. Also plotted are a few typical errors with the same symbols and colors. In short exposure observations (toward the left side of the figure), the number of detected sources is small (typically less than 10 in 10ks observations) and hence there is a large scatter with a correspondingly large error bar. On the other hand, in relatively long exposure observations (toward the right side of the figure), as the number of detected sources increases (typically 30-40 sources in 100ks observations), the error is small and hence the test result is more significant. The number of detected sources is consistent with the expected number. Among a dozen data points with exposure time ~100 ks, only one (obsid=536 in the H band) has a deviation larger (but only slightly) than the 1 σ error. Also marked in the figure are those data points determined in the two observations (shallow ACIS-S and much deeper ACIS-I observations) of the 3C295 field, where Cappi et al. (2001) reported the existence of an overdensity. The deeper observation clearly excludes this possibility. Delia et al. (2003, in preparation) found the same result with the deeper ACIS-I observations, i.e., no excess in this field of view, but also reported a smaller-scale, asymmetric distribution of X-ray sources. We will search for a scale-dependent density variation with the full ChaMP data set. We conclude that there is no indication of significant field-to-field variation in cosmic background source densities, within the statistics and over the flux levels included in our sample.

There have been some debates on whether the results from the CDF-N (Brandt et al. 2001) and CDF-S (Giacconi et al. 2001) are consistent with each other. Recently Rosati et al. (2002) reported that they are consistent within 2σ at the faint end (also consistent with our result), whereas the CDF-S contains a significantly lower number of sources at bright fluxes. We note that the source density in the CDF-S (N ~ 680 ± 150 deg$^{-2}$; Rosati et al. 2002) is consistent within ~1 σ with that in the CDF-N (N ~ 930 ± 300 deg$^{-2}$; Brandt et al. 2001) at $f_x=10^{-15}$ erg sec$^{-1}$ cm$^{-2}$ (in the S band), where our data overlap with the CDF-N data (see Figure 2). At bright fluxes, there are only a small number of sources in both CDFs. At $f_x$ ~ 8 x 10$^{-14}$ erg sec$^{-1}$ cm$^{-2}$ (in the S band), where the discrepancy appears largest, there are only ~5 real sources (with N ~ 70 deg$^{-2}$) in the CDF-S. At the same flux limit, our sample has ~200 sources, hence carrying much better statistics and providing a robust determination of the source density. Recently Yang et al. (2003) reported a field-to-field variation (mostly in the H-band) from the contiguous Lockman Hole NW covering 0.33 deg$^2$, while the Log(N)-Log(S) from the entire field is consistent with that of the CDF-N. However, it is not clear why the clustering is only significant in the H-band, but not in the S-band where the statistics are higher (~3 times more sources in the S-band). This requires further confirmation.

To test whether there is any systematic difference between fields with high and low galactic $N_H$ (or equivalently high and low galactic latitude), we have again divided our sample into 2 groups with $N_H$ higher or lower than $3 \times 10^{20}$ cm$^{-2}$. Again, we do not see any systematic difference (note that we applied ECF as a function of $N_H$.)

4. X-ray colors

4.1. Constructing X-ray colors

X-ray colors can be used to determine average spectral properties of a given sample of sources, which are not sufficiently bright to allow individual spectral fitting. We have developed two X-ray colors as defined in Table 1 to determine intrinsic spectral hardness and absorption. These colors are defined according to the convention used in optical colors – a large number in X-ray colors means a red (i.e., soft) spectrum. As in section 3, we have started with those sources of good quality (i.e., 3177 sources), but further limited to only sources with more than 50 net counts in the B band in CCDID=0-3 in ACIS-I and CCDID=6-7 in ACIS-S observations. We note that we include those sources which are not detected in all energy bands (they will have upper/lower limits in X-ray colors). Figure 5 shows the distribution of the resulting 620 sources in the X-ray color-color plane. All the X-ray sources used in this study are included in the first ChaMP catalog (Paper I). Sources from FI chips are plotted as green circles and sources from BI chips as blue stars. Also plotted are error bars representative for sources with 50 and 100 net counts with a typical spectrum ($\Gamma_{ph}$=1.7 and $N_H = 3 \times 10^{20}$ cm$^{-2}$). Note that those sources detected only in one band are marked with triangles pointing toward the appropriate direction in the color-color plane, i.e., at the top (no detection in H-band) or at the far-left (no detection in $S_1$-band).

The grid in the X-ray color-color plot (Figure 5) indicates the predicted locations of local sources at z=0 with various photon indices ($\Gamma_{ph}$ ranging from 0 to 4) and absorption column densities ($N_H$ ranging from $10^{20}$ to $10^{22}$ cm$^{-2}$). Note that C21 depends mostly on absorption (which changes X-ray colors horizontally), whereas C32 depends on intrinsic spectral hardness (which changes vertically). Because ACIS-S is more sensitive in the lower energies (E < 1 keV, or mostly in our $S_1$ band) than ACIS-I, the QE difference in two CCDs affects mostly C21 and the ACIS-S grid shifts to the right. The model prediction appropriate for ACIS-I is in green while that for ACIS-S is in blue, applying the same colors as the sources. As described in section 3, the model grid varies as a function of observation date, because of the ACIS QE degradation. X-ray colors of a typical source with $\Gamma_{ph}$=1.7 and $N_H = 3 \times 10^{20}$ cm$^{-2}$ which would have obtained in the earliest and latest observations (spanning about 20 months) differ by ~0.15 in C21 (much smaller in C32, ~0.02). This amount of ΔC21 is comparable to the error of typical sources with 100 counts (see Figure 5). The grids plotted in Figure 5 were made with obsid=00615 (for ACIS-S) and 00926 (for ACIS-I), which were observed near the mid-point of the 20 month span of our sample on July 10-11, 2000 (see Table 1 in Paper I). The X-ray colors are not corrected for vignetting, because the difference in the effective area at the off-axis angle of 10' between energies of 1.496 and 4.510 keV is only ~5% (see Chandra POG), or ~0.01 in X-ray colors, which is much smaller than the typical errors.

From Figure 5, it is clear that sources are concentrated at the location of typical broad-line AGNs with small intrinsic absorption, i.e., around $\Gamma_{ph}$=1.5-2 and $N_H$ = a few $\times 10^{20}$ cm$^{-2}$. About 70% of the sources in Figure 5 are, within statistical errors, consistent with being a typical broad-line AGN. However, there are a considerable number of outliers, which could consist of rare, interesting sources such as highly absorbed, very soft, or very hard sources. For example, the source at the bottom of Figure 5 (with C32 = -1.06), CXOMP J114147.9-660604, is a very hard source with almost all X-rays falling in the H band (note that C21 is very uncertain because of the small number of counts in E < 2.5 keV). This corresponds to $\Gamma_{ph} \leq -2$ if a significant amount of intrinsic absorption is absent. On the other hand, if this is due to absorption, $N_H \geq 10^{23}$ cm$^{-2}$ is required for a typical power-law spectrum with $\Gamma_{ph}$=1.5-2. As no sources with $\Gamma_{ph} \leq -2$ are known to exist, this source is likely a heavily absorbed AGN. Optical imaging data (Green et al. 2003; Optical spectroscopy will be obtained in 2003 Summer) also imply that it is optically a very red object, with $g' - i' > 4.8$ (or $g' - r' > 3.4$). With $r'$ =21.9 mag, its X-ray to optical flux ratio is $f_x/f_r \sim 1.0$, which is typical for AGNs (see Figure 7 and 8 below). There are several very soft sources at the top of Figure 5 (C32 ≥ 1.4), corresponding to $\Gamma_{ph} \geq 4$. Some of them could be X-ray stars (see Figure 9 in section 5). Also present are several absorbed sources at the far left of Figure 5, corresponding to absorption by $N_H \geq 10^{22}$ cm$^{-2}$. We expect to collect an order of magnitude more of these unusual sources in the full ChaMP sample. As follow-up optical imaging and spectroscopic observations progress (Green et al. 2003), we will be able to determine the nature of these peculiar sources.

4.2. X-ray color variation

As the XRB spectrum is harder than that of typical quasars (e.g., Gursky and Schwartz 1977), it has been suggested that the average source spectra would be harder with decreasing source fluxes (e.g., Comastri et al. 1995; Gilli et al. 1999). Recent observational data obtained by Chandra and XMM also support the spectral hardening (e.g., Hasinger et al. 2001; Alexander et al. 2002). They could consist of type 2 quasars (Webster et al. 1995) or red quasars (e.g., Kim and Elvis, 1999; White et al 2003). It is very important to confirm or reject this trend. Another equally interesting question is whether the hardening is caused by reducing soft X-rays due to the increasing amount of absorption or by enhancing hard X-rays due to intrinsic hardening. We can test this hypothesis by comparing the average X-ray colors of sources with different X-ray fluxes. Because the absorption affects C21 more than C32, we do not expect a similar trend in C32 as in C21 if the hardening is by the variable absorption of similar intrinsic spectra. On the other hand, if C32 exhibits a similar trend, i.e., harder colors with decreasing fluxes, then fainter sources could be intrinsically hard in X-rays.

In the top panel of Figure 6, C21 and C32 are plotted against count rates. There is a clear trend in C21 (Figure 6-a) such that sources with hard C21 colors appear to be X-ray faint, while X-ray bright sources have all soft C21 colors. In other words, the low right corner of Figure 6-a is almost empty. We ran a K-S (Kolmogorov-Smirnov) test to two sub-groups with X-ray B-band count rate higher or lower than 5 count ksec$^{-1}$ (the vertical lines in the top panel indicate this count rate). The probability that the X-ray bright and faint samples originate from the same parent population is $\sim 10^{-5}$, indicating that the trend in the C21 color is statistically significant. To further illustrate, we separately plotted them in the color-color diagram (the bottom panel of Figure 6). The location for $N_H > 3 \times 10^{21}$ cm$^{-2}$ (i.e., the left side of the dashed line between two solid lines indicating $N_H = 10^{21}$ and $10^{22}$) is almost empty for X-ray bright sources (Figure 6-c), whereas the same space is occupied by many X-ray faint sources (Figure 6-d). This is in good agreement with the expectation from the hard XRB spectra in that the faint sources (which could consist of a large part of XRB but are mostly undetected so far) are indeed more absorbed, hence harder.

In C32, the trend is much less clear. In Figure 6-a, the hard X-ray sources with C21 < -0.5 found in FI chips (or C21 < -0.3 in BI chips, considering the higher response of BI at lower energies) are color-coded with red. The same objects are also marked in red in Figure 6-b where C32 is plotted against the X-ray count rate. While the hard X-ray sources are clearly distinguishable in C21, they are all mixed with soft X-ray sources in C32. Likewise, the faint sources found in the left side of the color-color plane in Figure 6-c ($N_H > 3 \times 10^{21}$ cm$^{-2}$) have a similar distribution across the $\Gamma_{ph}$ grids to those sources on the right side ($N_H < 3 \times 10^{21}$ cm$^{-2}$), i.e., they are concentrated around the location with $\Gamma_{ph}$=1.5-2 with some scatter, but without a clear preference toward higher or lower $\Gamma_{ph}$. The K-S test resulted in a much higher probability of the two datasets being drawn from the same population (~$10^{-2}$) than that for C21. The absence of any systematic trend in C32 is consistent with the results in Green et al. (2003), where a single hardness ratio was used (i.e., compared counts in the S (= $S_1$+$S_2$) band and in the H band). This implies that the spectral hardening of faint sources is mostly due to absorption which affects the lowest energies (thus the C21 color), and the intrinsic spectral shape on average does not change with decreasing X-ray flux.

We note that with the chosen energy bands (Section 2) and the given sample selection (above in this section), we have upper limits in C21 (for sources detected in the $S_2$ band, but not detected in the $S_1$ band), but no lower limits. On the other hand, we have only lower limits in C32 (for sources detected in the $S_2$ band, but not detected in the H band), but no upper limits. Including the upper/lower limits would add more *hard* sources (mainly to the X-ray faint sub-group) in the C21-$f_x$ relationship, and more *soft* sources in the C32-$f_x$ relationship. In both cases, this would slightly enhance (but not reduce) our conclusions concerning the presence/absence of the correlation of C21/C32 with $f_x$.

5. Cross-correlation with other catalogs

With Chandra's superb spatial resolution, the typical error in absolute position is 1″ or less (see Section 3.4 in Paper I and http://cxc.harvard.edu/cal/ASPECT/celmon) and the relative error of **wavdetect** determined positions is less than 1-2″ within $D_{off-axis} < 400″$ (see Section 5 in Paper I). This makes an error box at least an order of magnitude smaller than those typical of previous missions (e.g., ROSAT WGA catalog; White et al. 1995). Cross-correlation with existing catalogs in other wavelengths are now substantially more reliable without ambiguity.

As in section 3 and 4, we have selected only sources of good quality (i.e., 3177 sources), but further limited to X-ray sources with a good positional accuracy by applying (1) net count in the B band > 10 and (2) $D_{off-axis} < 400″$ regardless of ACIS-I or ACIS-S observations. Then, we have cross-correlated our data with optical (US Naval Observatory, USNO A2.0; Monet et al. 1998), near-IR (Two Micron All Sky Survey, 2MASS; Cutri et al. 2000) and radio (NVSS; Condon et al. 1998 and FIRST; White et al. 1997) catalogs, with a 3″ search radius. We have found 207, 75, and 16 sources with USNO, 2MASS and NVSS+FIRST counterparts, respectively. Here we will concentrate on the correlation with the USNO catalog and the other results (and with a full ChaMP dataset) will be presented in a separate paper.

In figure 7, 207 sources with USNO optical counterparts are plotted in the $f_x$ – R plane. The X-ray flux is determined in 0.5-2.0 keV and the X-ray to optical flux ratio is given by log $f_x/f_R$ = log $f_x$ +

5.57 + R/2.5 [taken from Maccacaro et al. 1988, assuming B-R=1.0 mag and V=(B+R)/2.0]. Also plotted are diagonal lines indicating $f_x/f_R$ = 0.01 – 10. As determined in EMSS (Maccacaro et al. 1988) and also confirmed by the ChaMP (Green et al. 2003), the upper-right corner (i.e., $f_x > 0.1 f_R$) is mainly occupied by quasars whereas the lower-left corner (i.e., $f_x < 0.1 f_R$) is for galactic stars and galaxies (including both narrow emission line galaxies and absorption line galaxies). This figure can be directly compared with ChaMP optical results (e.g., Green et al. 2003), where the CCD-determined optical magnitude goes much deeper ($r' \sim 24$ mag). Note that those with USNO IDs are only ~10% of the sources cross-correlated in this section. Since stars and galaxies are brighter in R for a given $f_x$ than are quasars, the relatively bright optical limit of USNO gives a biased fraction of stars and galaxies, which will change as we go fainter in R.

In Figure 8, optical colors are plotted against X-ray to optical flux ratio. The X-ray and R-band fluxes and their ratios are determined as in Figure 7. Again this figure can be directly compared with a similar plot in Green et al. (2003). The vertical dashed line corresponds to $f_x = 0.1 f_R$, so the right side of that line is the location for quasars. As seen in Green et al. (2003), the typical blue quasars have B-R < 1 (i.e., below the horizontal dashed line). In Figure 8, we can see a concentration of typical blue quasars in the lower-right region, defined by $f_x > 0.1 f_R$ and B-R < 1.0 mag. The left side of the vertical lines is the location for galactic stars and galaxies, while absorbed, red quasars or type 2 quasars would fall in the upper-right corner, defined by $f_x > 0.1 f_R$ and B-R > 1.0 mag. So the $f_x/f_R$ – B-R plane can be divided into 3 sub-regions, hosting blue quasars (lower-right), red quasars (upper-right) and stars/galaxies (left). Using X-ray colors, we determine average X-ray spectral properties of sources falling in these 3 sub-regions. X-ray color-color diagrams for each sub-sample (with net counts > 50 as in Figure 5) are shown in Figure 9. The blue quasars and stars/galaxies occupy different regions of the X-ray color-color diagram. The blue quasars are clearly concentrated in the region for $\Gamma_{ph}$=1.5-2 and $N_H$= a few x $10^{20}$ cm$^{-2}$. On the other hand, stars/galaxies are widely spread out, particularly occupying a region with higher (or softer) C32 color ≥ 1.0, which corresponds to $\Gamma_{ph} \geq 3$. We note that for a thermal gas model the region with C32 ≥ 1.0 corresponds to kT ≤ 1.5 keV, which is typical for galaxies and groups of galaxies. We do not have many objects in the region of red quasars (with more than 50 X-ray net counts) in our sample to draw any characteristic signature using the X-ray colors. However, they appear to spread out, contrary to the concentrated distribution of blue quasars. The full ChaMP sample with deep optical imaging data supplemented will allow us to revisit this important issue of the existence, fraction and properties of red and/or absorbed quasars.

Among ~1900 sources with no optical counterpart in the USNO catalog, more than a half of them could have $f_x > 0.1 f_R$, being consistent with a typical quasar. We note that 9 sources satisfy the 'blank' source criteria ($f_x / f_V > 10$), as applied in Cagnoni et al. (2002). For these 9 sources, we do not see any extra X-ray absorption, based on X-ray colors, consistent with Cagnoni et al. (2002). None of them is extended, but 1 of them could be variable. Among the sources (i.e., 3177 sources) found in CCDID=0-3 in ACIS-I and CCDID=6-7 in ACIS-S observations, we have identified 21 extended sources and 92 variable sources. They will be presented in a separate paper along with those 'blank' sources. The full ChaMP follow-up observations will provide statistical samples of a variety of intriguing objects with high quality, unambiguous optical counterparts and an order of magnitude more in quantity.

## 6. Conclusion

Analyzing ~1000 near on-axis, bright X-ray sources from the initial sample of uniformly reduced 62 Chandra observations, we conclude that:

(1) Our Log(N)-Log(S) produced by the initial ChaMP sample is consistent with those seen by the Chandra deep surveys and the previous missions. In particular, the H-band Log(N)-Log(S) fills the gap between the CDF and ASCA data. The best fit slopes for the S-band differential Log(N)-Log(S) are $\beta_{bright} = 2.2 \pm 0.2$ and $\beta_{faint} = 1.4 \pm 0.3$ with $S_{break} = 6 (\pm 2) \times 10^{-15}$ erg sec$^{-1}$ cm$^{-2}$, while $\beta = 2.3 \pm 0.1$ for the H-band differential Log(N)-Log(S).

(2) There is no statistically significant difference between the Log(N)-Log(S) produced by fields containing clusters of galaxies and fields without clusters. Also there is no statistically significant (within ~ 1 $\sigma$) field-to-field variation in number densities of cosmic X-ray background sources in the flux range of $f_X \sim 10^{-15} - 10^{-13}$ erg sec$^{-1}$.

(3) The majority (~ 70%) of X-ray sources exhibit X-ray colors, typical of unabsorbed ($N_H$ = a few $\times 10^{20}$ cm$^{-2}$) quasars (a photon index ~ 1.7). There also exist heavily absorbed, or very soft/hard sources.

(4) On average the X-ray color becomes harder as the count rate decreases. It appears that this is mainly because of absorption rather than intrinsic spectral hardening because the hardening occurs in the soft color, but does not exist in the hard color.

(5) Cross-correlating with optical catalogs, we have divided X-ray sources into 3 groups: blue quasars, red quasars, and stars and normal galaxies. In the X-ray color-color plot, the blue quasars are well concentrated near the location of a typical unabsorbed power-law. On the other hand, the stars/galaxies exhibit a wide spread, and occupy a location with kT < 1.5 keV. Only a small number of red quasars are found, but from the full ChaMP sample we expect a considerable number of such sources which will allow us to test whether they are really obscured either or both in the optical and X-ray band.

This work has been supported by CXC archival research grant AR2-3009X. We acknowledge support through NASA Contract NAS8-39073 (CXC). We thank the CXC DS and SDS teams for their supports in pipeline processing and data analysis.

Yang et al., 2003, astro-ph/0302137

Figure Captions

Figure 1. Log(N)–Log(S) in the soft energy band (0.5-2 keV; left panel) and in the hard band (2.0-8.0 keV; right panel), measured separately in ACIS-I (ccdid=0-3; top panel) and ACIS-S (ccdid=7; bottom panel). The solid histogram and open circles with error bars are our unbinned and binned data, respectively. Also plotted for comparison are results from Hubble Deep Field-North and previous missions (ROSAT in the S band and ASCA data in the H band).

Figure 2. Same as Figure 1, but with all sources detected in ACIS-I and ACIS-S. The dashed line in the right panel indicates the results of the ASCA fluctuation analysis.

Figure 3. Same as Figure 1, but made separately with sources found in the fields with (top panel) and without (bottom panel) an X-ray cluster.

Figure 4. The number of detected sources compared with the number of expected sources, based on our fitted LogN – LogS. Filled (open) circles indicate sources detected in the S (H) band of 4 ACIS-I CCDs within $400''$ from the aim point. Similarly, squares are for the ACIS-S CCD (S3). Typical errors are plotted with the same symbol and color as data points. Also the data obtained in the 3C295 field are marked by arrows.

Figure 5. X-ray color-color plot. X-ray colors, C21 and C32, are defined as in Table 1. The sources detected with more than 50 counts in front-illuminated (FI) and back-illuminated (BI) chips are marked by green circles and blue stars, respectively. The upper limits of X-ray colors for sources detected only in one band are marked by triangles on the top and on the far-left. The green (FI) and blue (BI) grids indicate the location of predicted X-ray colors of power-law spectra with specified parameters. The two error bars on the lower-right corner represent typical errors for sources with 50 and 100 net counts.

Figure 6. (top panel; a and b) X-ray colors as a function of X-ray count rate in B-band. The color-code is the same as in Figure 5, except that absorbed sources with C21 < -0.5 for FI (or C21 < -0.3 for BI) are marked by red circles/stars. (bottom panel; c and d) The same as in Figure 5, except sources with net count rate in B-band (c) higher and (d) lower than 5 count ksec$^{-1}$ (vertical dashed lines in the top panel) are plotted separately.

Figure 7. X-ray sources with USNO optical counterparts. The X-ray flux is determined in the 0.5-2.0 keV band and the X-ray to optical flux ratio is given by log $f_x/f_R$ = log $f_x$ + 5.57 + R/2.5. The upper-right corner (i.e., $f_x$ > 0.1 $f_R$ ; indicated by the green solid line) is mainly occupied by quasars whereas the lower-left corner (i.e., $f_x$ < 0.1 $f_R$) is primarily local stars and galaxies, as determined in EMSS and ChaMP (see also Green et al. 2003).

Figure 8. Optical colors are plotted against X-ray to optical flux ratio. The X-ray and R-band fluxes are determined as in Figure 7. The vertical dotted line corresponds to $f_x$ = 0.1 $f_R$, so the right side of that line is the location for quasars. The typical blue quasars have B-R < 1 (i.e., below the horizontal dotted line) whereas the absorbed, red quasars would fall above the line with B-R > 1, as determined in Kim and Elvis (1999) and also ChaMP (Green et al. 2003).

Figure 9. X-ray color-color diagram. Same as in Figure 5. (a) all X-ray sources with USNO optical counterparts, (b) sources (possibly red quasars) in the upper right corner of Figure 8 (i.e., B-R > 1 and $f_x > 0.1\ f_R$), (c) sources (stars/galaxies) in the left side of Figure 8, and (d) sources (blue quasars) in the lower-right corner of Figure 8.

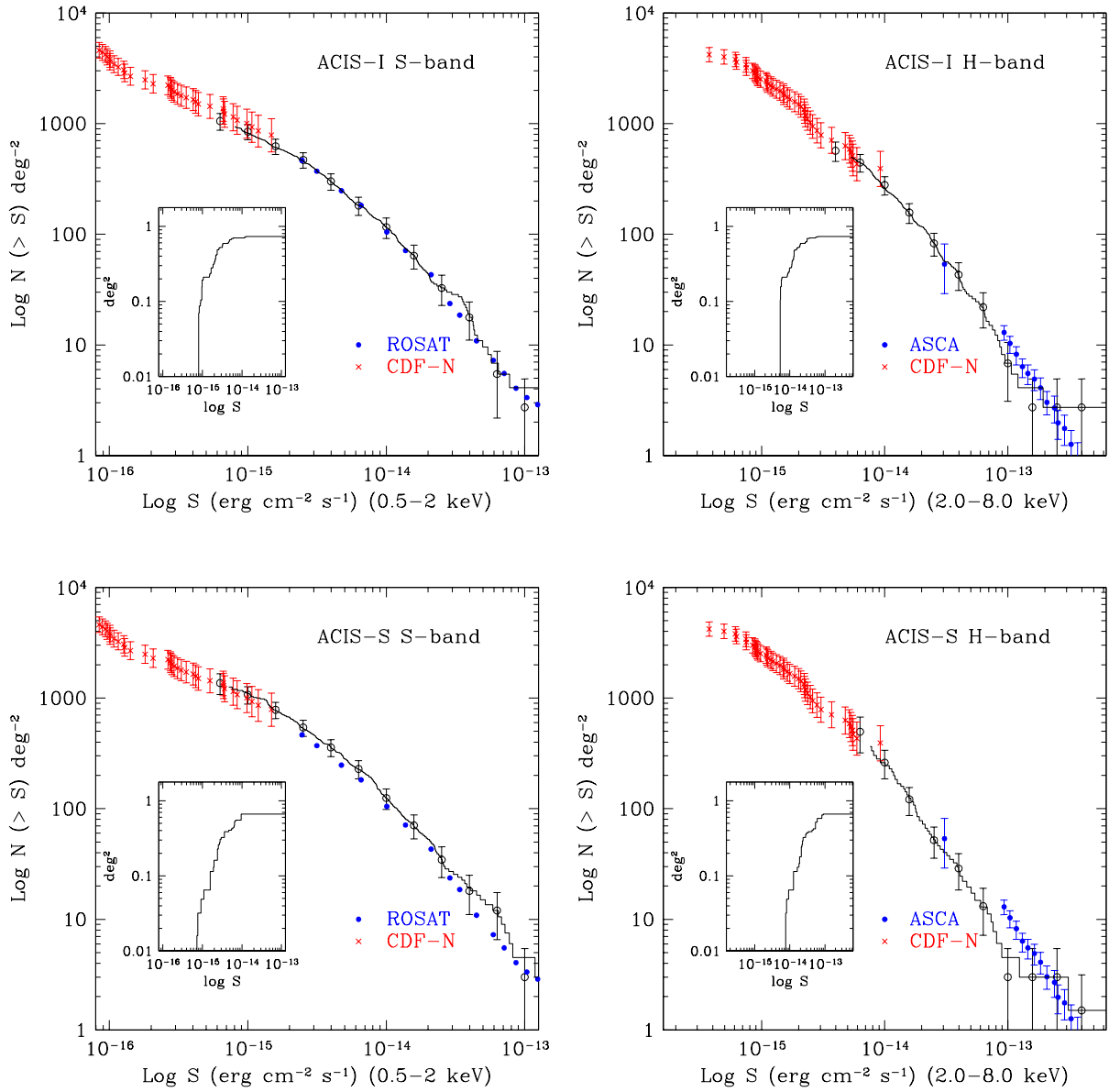

Figure 1: Log(N)-Log(S) in the soft energy band (0.5-2 keV; left panel) and in the hard band (2.0-8.0 keV; right panel), measured separately in ACIS-I (ccdid=0-3; top panel) and ACIS-S (ccdid=7; bottom panel). Also plotted for comparison are results from Hubble Deep Field-North and previous missions (ROSAT in the S band and ASCA data in the H band).

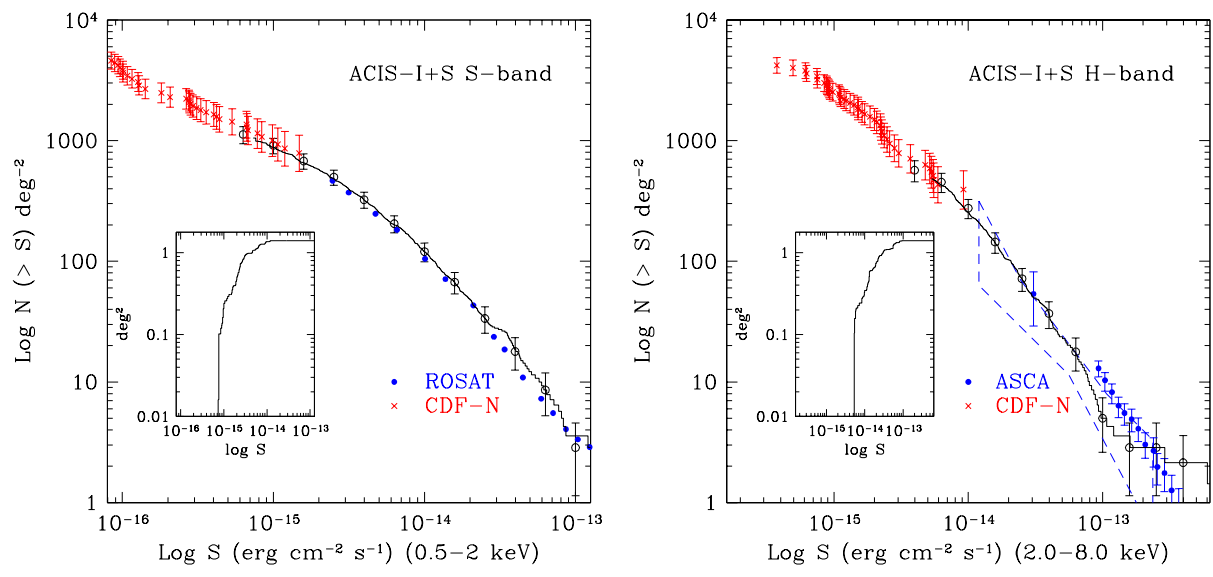

Figure 2: Same as Figure 1, but with all sources detected in ACIS-I and ACIS-S. The dashed line in the right panel indicates the results of the ASCA fluctuation analysis.

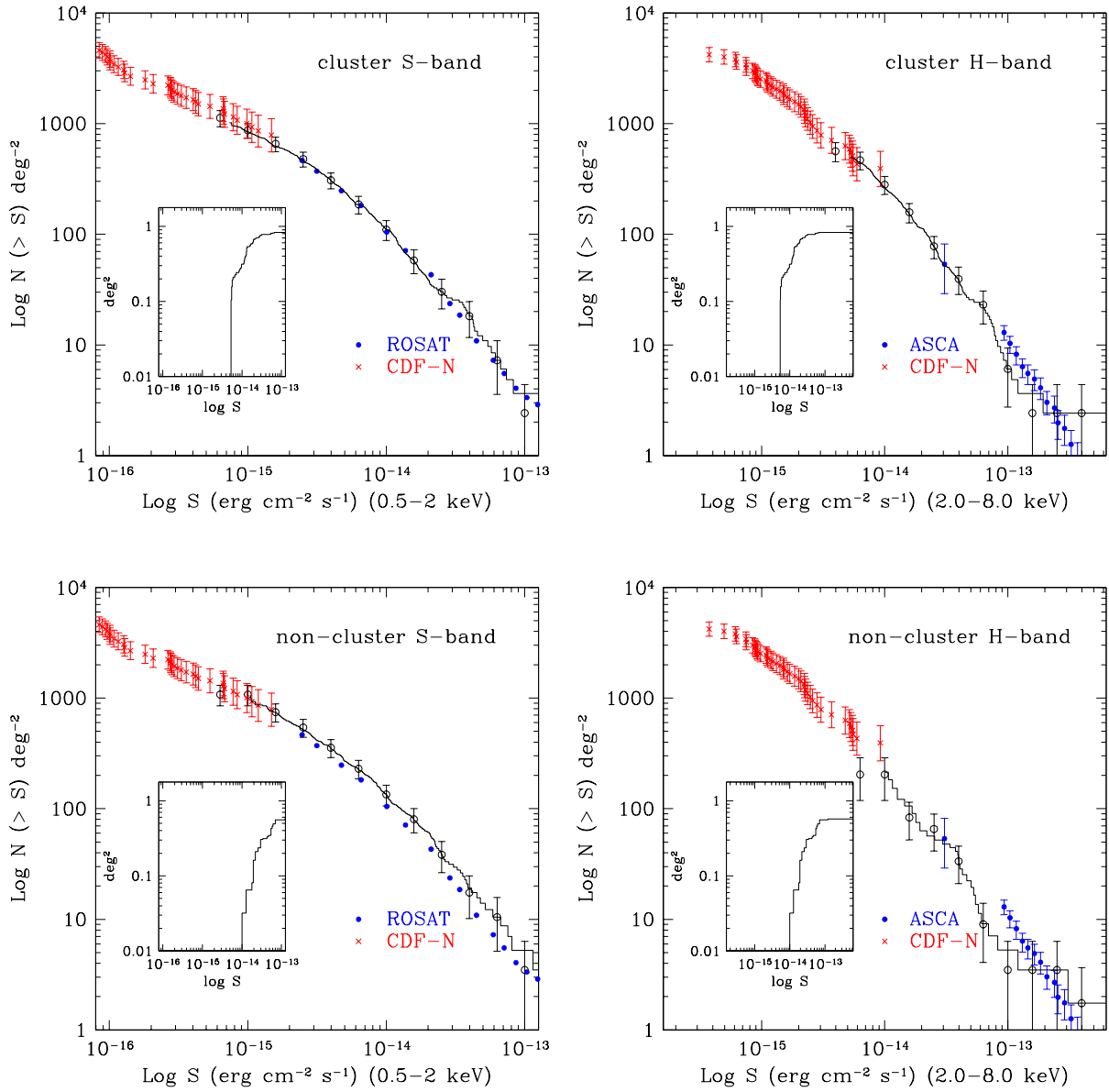

Figure 3: Same as Figure 1, but made separately with sources found in the fields with (top panel) and without (bottom panel) an X-ray cluster.

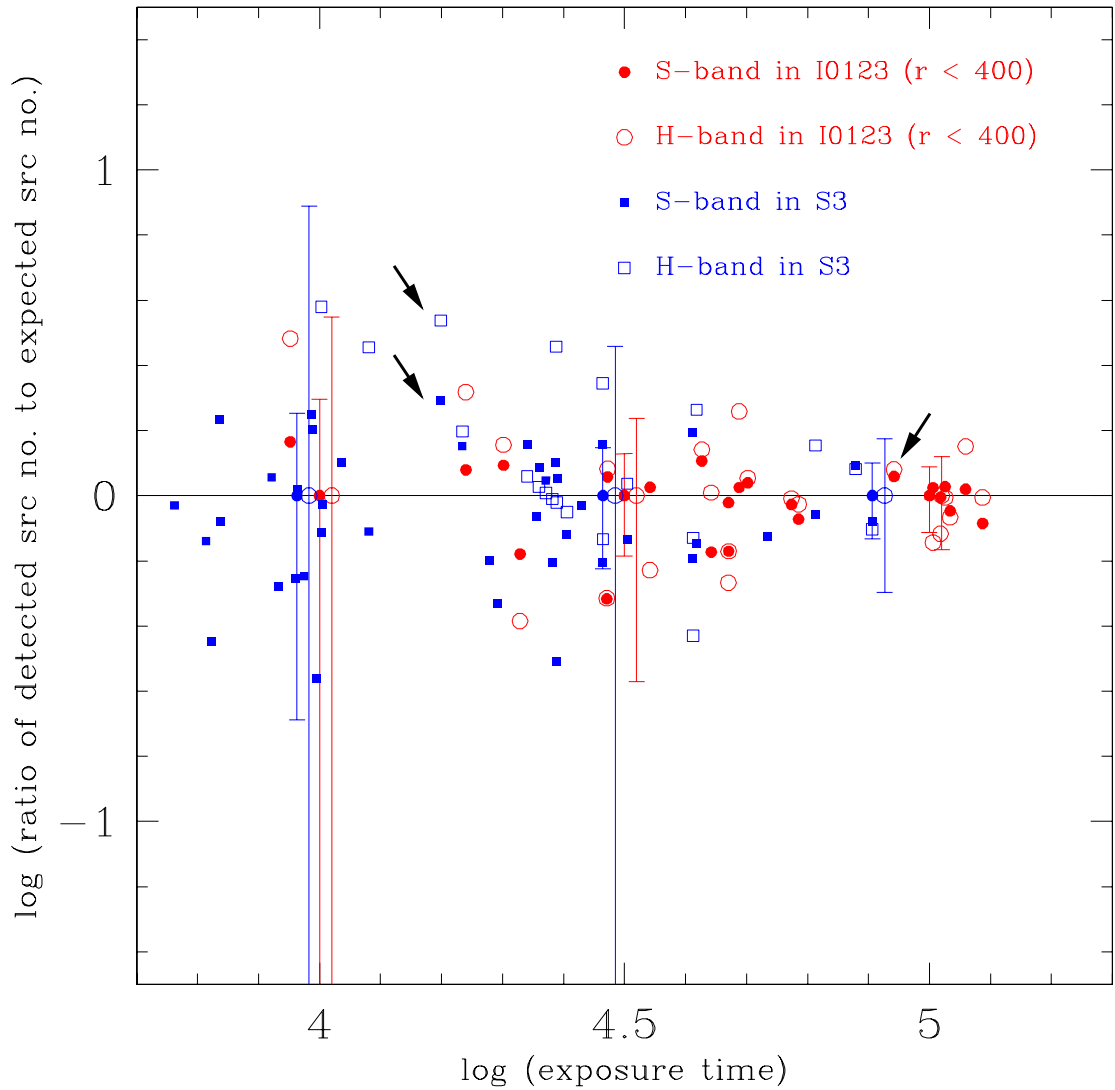

Figure 4: The number of detected sources compared with the number of expected sources, based on our fitted LogN - LogS. Filled (open) circles indicate sources detected in the S (H) band of 4 ACIS-I CCDs within 400" from the aim point. Similarly, squares are for the ACIS-S CCD (S3). Typical errors are plotted with the same symbol and color as data points. Also the data obtained in the 3C295 field are marked by arrows.

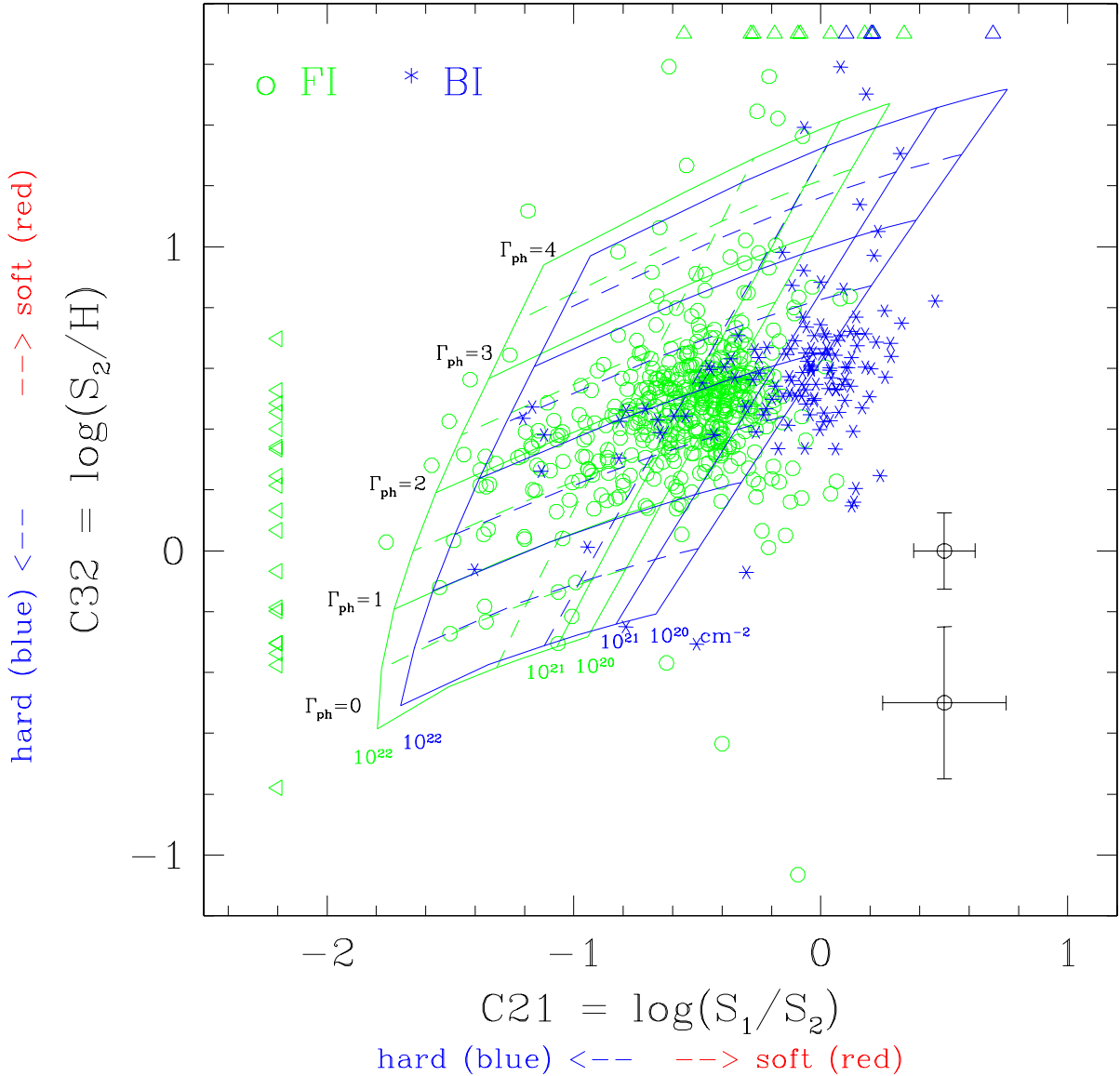

Figure 5: X-ray color-color plot. X-ray colors, C21 and C32, are defined as in Table 1. The sources detected with more than 50 counts in front-illuminated (FI) and back-illuminated (BI) chips are marked by green circles and blue stars, respectively. The upper limits of X-ray colors for sources detected only in one band are marked by triangles on the top and on the far-left. The green (FI) and blue (BI) grids indicate the location of predicted X-ray colors of power-law spectra with specified parameters. The two error bars on the lower-right corner represent typical errors for sources with 50 and 100 net counts.

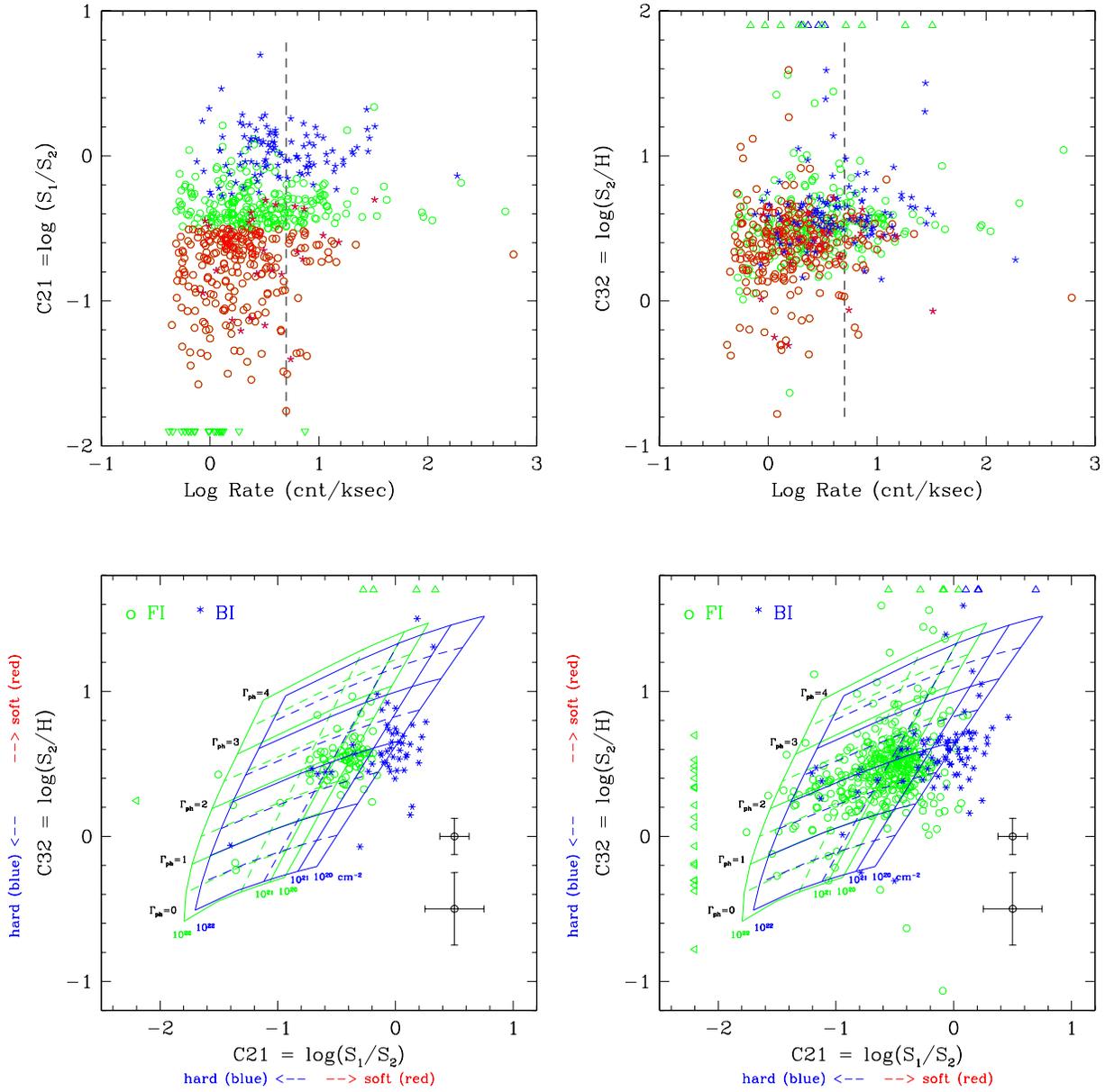

Figure 6: (top panel; a and b) X-ray colors as a function of X-ray count rate in B-band. The color-code is the same as in Figure 5, except that absorbed sources with C21 < -0.5 for FI (or C21 < -0.3 for BI) are marked by red circles/stars. (bottom panel; c and d) The same as in Figure 5, except sources with net count rate in B-band (c) higher and (d) lower than 5 count ksec$^{-1}$ (vertical dashed lines in the top panel) are plotted separately.

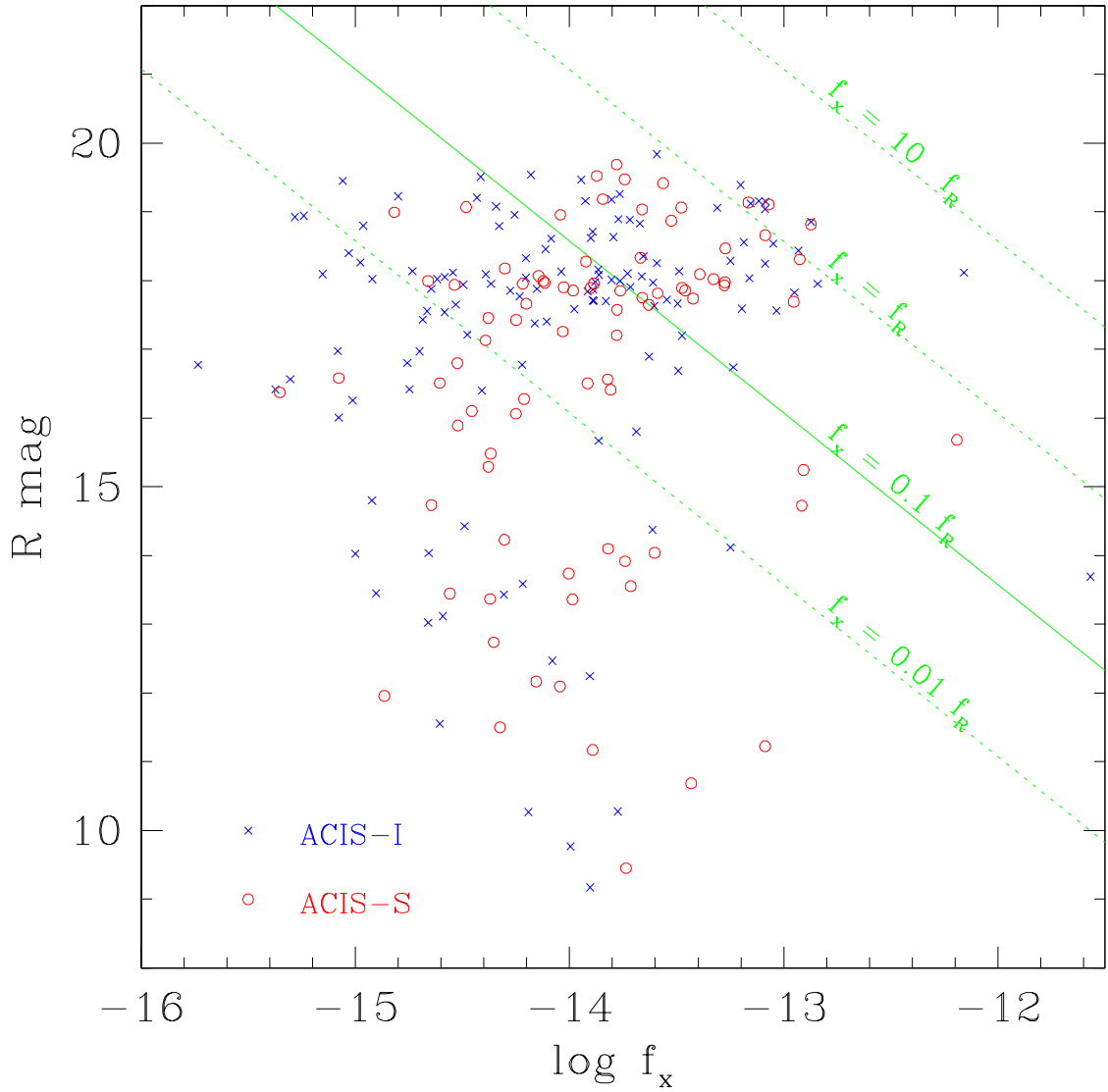

Figure 7: X-ray sources with USNO optical counterparts. The X-ray flux is determined in the 0.5-2.0 keV band and the X-ray to optical flux ratio is given by log fx/fR = log fx + 5.57 + R/2.5. The upper-right corner (i.e., fx > 0.1 fR ; indicated by the green solid line) is mainly occupied by quasars whereas the lower-left corner (i.e., fx < 0.1 fR) is primarily local stars and galaxies, as determined in EMSS and ChaMP (see also Green et al. 2003).

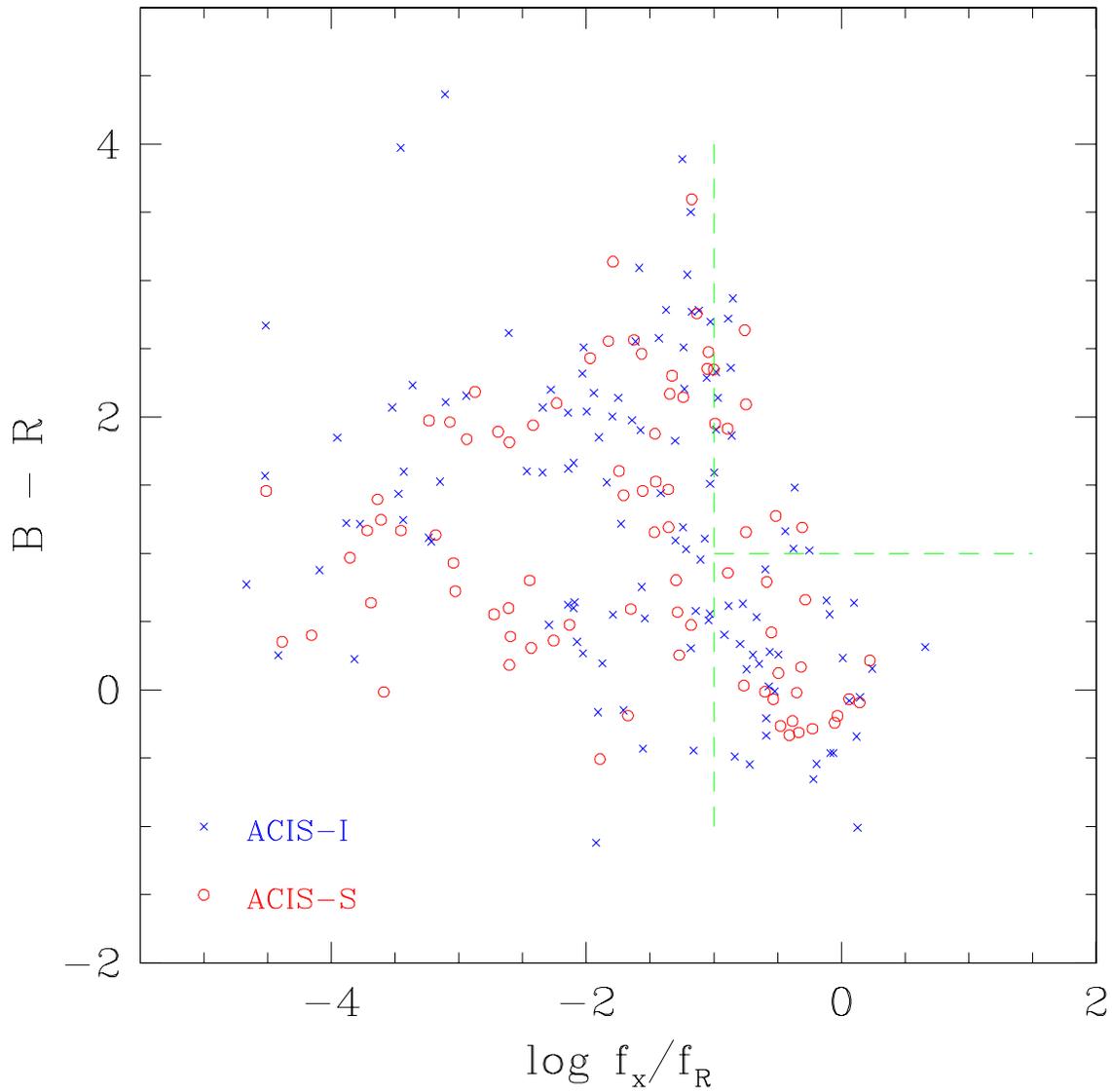

Figure 8: Optical colors are plotted against X-ray to optical flux ratio. The X-ray and R-band fluxes are determined as in Figure 7. The vertical dotted line corresponds to $f_x = 0.1 f_R$, so the right side of that line is the location for quasars. The typical blue quasars have B-R < 1 (i.e., below the horizontal dotted line) whereas the absorbed, red quasars would fall above the line with B-R > 1, as determined in Kim and Elvis (1999) and also ChaMP (Green et al. 2003).